\documentclass[aip,graphicx,reprint,nofootinbib, doi=false]{revtex4-1}
\usepackage{amsmath}
\usepackage{graphicx}
\usepackage{gensymb}
\usepackage{commath}
\usepackage{physics}

\usepackage{svg}
\usepackage[export]{adjustbox}

\graphicspath{ {./images/} }

\begin{document}


\title{Design and fabrication of a micro-ion trap with a 3D-printed loading zone for improved hot-ion capture}



\author{Sayan Patra}
\email{patra2@llnl.gov}
\affiliation{Lawrence Livermore National Laboratory, 7000 East Ave., Livermore, CA 94550, USA}
\author{Abhinav Parakh}
\affiliation{Lawrence Livermore National Laboratory, 7000 East Ave., Livermore, CA 94550, USA}
\author{Xiaoxing Xia}
\affiliation{Lawrence Livermore National Laboratory, 7000 East Ave., Livermore, CA 94550, USA}

\author{Juergen Biener}
\affiliation{Lawrence Livermore National Laboratory, 7000 East Ave., Livermore, CA 94550, USA}
\author{Hartmut H\"{a}ffner}
\affiliation{Department of Physics, University of California, Berkeley, Berkeley, California 94720, USA}
\affiliation{Challenge Institute for Quantum Computation, University of California, Berkeley, Berkeley, California 94720, USA}
\author{Kristin M. Beck}
\email{beck37@llnl.gov}
\affiliation{Lawrence Livermore National Laboratory, 7000 East Ave., Livermore, CA 94550, USA}

\date{\today}

\begin{abstract}

\noindent We leverage recent advances in 3D-printing technology to design and fabricate a micro-ion trap with a spatially distinct loading zone for more efficient loading of ions from effusive thermal ovens. The design reduces the Mathieu-$q$ parameter in the loading zone by increasing the ion-electrode separation $r_0$, thereby potentially facilitating more effective laser cooling of hot ions. This circumvents the temporary thermal instability that arises when the rf potential is reduced during ion loading, a common practice to enable efficient laser cooling of hot ions. Simulations predict that expanding $r_0$ maintains a high trapped ion fraction from a simulated thermal source across a wide range of Mathieu-$q$ parameters. We demonstrate the manufacturability of this design by 3D-printing the rf rails of a four-rod ion trap and discuss the limitations imposed by state-of-the-art additive manufacturing techniques. We briefly compare hot-ion capture in the three-dimensional design presented here with that in a representative planar trap, illustrating one instance in which the former may be better for loading. The article concludes with an outlook for how this design may be incorporated into a quantum-CCD architecture to enhance ion loading and reduce associated experimental overheads.

\end{abstract}


\maketitle 

\section{Introduction}\label{intro}

\noindent A key component of a trapped-ion quantum computer is an ion trap that supports high motional frequencies, enabling high-fidelity quantum operations, while providing ample optical access for rapid state readout of trapped-ion qubits. To enable scalability, these traps are designed with small ion-electrode separations (about $100~\mu\textup{m}$), allowing ions to be confined with motional frequencies in the $3-5~\textup{MHz}$ range at moderate peak voltages of $100~\textup{V}$. Surface electrode ion traps~(SETs) were proposed\cite{Chiaverini_2005} and subsequently developed\cite{Pearson_2006, Stick_2006, Tanaka_2009, Amini_2010, Maunz_2016, Moses_2023} to satisfy these requirements. The central insight of this platform is that a two-dimensional array of electrodes provides sufficient field control and supports voltages high enough to capture, trap, and coherently transport multiple ions close to their motional ground state. This enables devices to be made using micro-machining or lithography with higher accuracy and precision, and improves repeatability by removing the tolerancing from manual assembly. However, the planar geometry is not without disadvantages. For similar dimensions and operating parameters, a planar ion trap generates a shallower and more strongly anharmonic potential well with a lower radial motional frequency than its three-dimensional counterpart\cite{Chiaverini_2005}. A shallower well lowers the capture probability of hot ions during loading, while anharmonicity enforces stricter motional stability criteria than a harmonic potential. Lower motional frequencies of the stored ions result in higher sensitivity to $1/f$ electric-field noise and slower transport operations such as ion swapping~\cite{Kaufmann_2017} and crystal reordering. An ideal hardware platform for a trapped-ion quantum computer should combine the deep and harmonic rf confinement of a three-dimensional Paul trap with the manufacturability and scalability of an SET.\\

\noindent In modern trap architectures aimed at scalability, the loading region is spatially separated from the experimental regions\cite{Revelle_phoenix_2020, Moses_2023}. This keeps the trap surfaces in experimental regions isolated from the metallic deposition caused by repeated ion loading, which results in excess electric-field noise and increased motional heating in ions trapped in the loading zones~\cite{Turchette_2000, Daniilidis_2011}. However, the rf confinement remains same in both the regions; a single rf feed delivers power to all trapping regions. A common practice to enable efficient ion loading in these traps is to temporarily reduce the delivered rf power. This leads to a transient thermal instability. As an alternative to this temporal protocol, we propose a spatial one in which the ion-electrode separation is increased in the loading zone while maintaining the requisite confinement in the experimental regions.\\

\noindent To realize such a geometry, we leverage two-photon polymerization direct laser writing (2PP-DLW). This additive manufacturing technique was recently employed to fabricate an operational ion trap by some of the authors of this work~\cite{Xu_2025}. Four rf blades were printed perpendicular to a surface to form a three-dimensional ion trap. Surface electrodes apply direct current (dc) potentials to compensate micromotion or move ions. The trap demonstrated state-of-the-art performance with operation close to the edge of the first zone of the stability diagram, long single-ion lifetimes, and a M{\o}lmer-S{\o}rensen gate with $0.978\pm0.012$ fidelity. In this article, we design, analyze, and fabricate our proposed geometry that is expected to improve the efficiency of loading ions in a 3D-printed four-rod trap.\\

\section{Motivation}\label{motivation}

\noindent A trapped-ion quantum computer typically consists of one or more strings of ions confined axially in an ion trap. The number of ions depends on architectural choices; common choices are two to four in the QCCD architecture~\cite{Delaney_2024}, and tens in a MUSIQC device~\cite{Chen_2024} or quantum simulator~\cite{Zhang_2017}. To ensure that a cold ion crystal forms a linear string, the radial confinement needs to be stronger than the axial confinement such that~\cite{Nagerl_2000} 
\begin{equation}\label{Eq:0}
	\omega_\textup{rad}/\omega_\textup{ax}>0.73N^{0.86},
\end{equation}
\noindent where $N$ is the number of ions in the string, and $\omega_\textup{rad}$ and $\omega_\textup{ax}$ are, respectively, the radial and axial secular frequencies of a trapped ion. Ion loading and quantum-gate performance impose competing requirements. Higher motional frequencies support faster and less error-prone gates and reduce sensitivity to the $1/f$  effect of electric-field noise \cite{Deslaurier_2006}. These radial secular frequencies are achieved by pushing the Mathieu-$q$ parameter $q$ to higher values, which requires a trade-off between decreasing the rf frequency $\Omega_\textup{T}$ and increasing the peak voltages $V_0$~(see Eq.~\ref{App:Eq3a}). While operating a trap in this regime is optimal for quantum operations, the high $q$ limits the trapping volume and results in inefficient ion loading~\cite{Gerlich_1992}.\\

\noindent Ions are typically produced by photoionization of neutral atoms, close to the minimum of the trapping potential generated by the electrodes. Efficient loading requires capturing and confining a large fraction of these ions. The trap depth, quantified for small $q$ by the pseudopotential well depth $U_\textup{pseudoPE}$, plays an important role in capturing hot ions. In planar traps, $U_\textup{pseudoPE}$ is comparable to the average kinetic energy of ions ($\sim0.1~\textup{eV}$) produced from a locally heated effusive oven, leading to poor capture probability. Capture is poorer for ions created from laser ablated atoms~\cite{Smith_2025, Greenberg_2024}, whose average kinetic energies are much greater than $0.1~\textup{eV}$, corresponding to temperatures far above $1000~\textup{K}$. Stable trajectories of the captured ions are essential for efficient loading. The potentials in surface electrode traps have significant higher-order multipolar contributions, which have more stringent criteria for stable ion motion compared to a pure quadrupolar potential, thereby degrading ion loading~\cite{Mathiesen_2021}. Loading inefficiency may be further exacerbated for architectures with backside loading on account of weaker confinement normal to the electrode plane~\cite{Revelle_phoenix_2020}. Reduced loading efficiency was already identified as an issue with SETs in Ref.~[\onlinecite{Chiaverini_2005}], where the authors proposed circumventing the issue by integrating a macroscopic three-dimensional ion trap\cite{Naegerl_1998, Berkeland_2002} as an ion reservoir.\\

\noindent Efficient ion trapping requires efficient laser cooling. Trapped hot ions have large motional amplitudes, which show up as frequency modulation on the Doppler-cooling transition and reduce cooling efficiency\cite{DeVoe_1989, Peik_1999, Wesenberg_2007, Epstein_2007, Sikorsky_2017}. During loading, multiple ions are trapped. The relative motion of the trapped ions leads to convoluted absorption profile of the Doppler-cooling laser, and imposes stringent parameters for efficient laser cooling~\cite{vanMourik_2022}. In an experiment, this is typically manifested as an ensemble of trapped ions forming an unstructured cloud ($T_\textup{ion}\gg10~\textup{mK}$) instead of a Coulomb crystal close to the Doppler cooling limit.\\

\noindent An efficient but operationally expensive solution to load ions in shallow and stiff traps is to use an ensemble of pre-cooled neutral atoms~\cite{Sage_2012, Johansen_2022}. Using cold neutrals enables better isotope selectivity, larger capture fraction, and faster site-selective reloading~\cite{Bruzewicz_2016} at the cost of significant additional infrastructure. A solution to improve capture and efficiently Doppler laser-cool hot ions from thermal sources would therefore be beneficial for loading ions.\\ 

\noindent Reducing the Mathieu-$q$ parameter of a trap has been observed to facilitate efficient laser cooling. Operationally, this entails a reduction in the applied rf voltage, leading to a weaker electric field. Consequently, an ion created away from the rf null in such a trap gains less energy from the field compared to a stiffer trap, thereby enabling faster laser cooling. Reducing $q$ also results in weaker micromotion sidebands, which lead to a simpler absorption profile of the Doppler-cooling laser. Cooling is further enhanced with an additional far detuned cooling beam that addresses hot ions with large Doppler shifts~\cite{vanMourik_2022}. After the ions are cooled to form crystals, the rf potential is increased to achieve the motional frequencies desired for coherent operations. This temporal change in rf potential induces transient thermal instability in the trap, leading to instabilities in the ions’ secular frequencies. These transients may increase operational overheads, either in the form of degraded fidelity of quantum operations, or wait time for the motional frequencies to stabilize. In a multi-zone device, this temporal protocol is not a scalable approach. Rf confinement is decreased in all trapping sites, so loading cannot occur while high fidelity quantum operations are being performed elsewhere on the chip.\\

\noindent The loading protocol described above may also be implemented spatially instead of temporally, which avoids the aforementioned thermal instability and scheduling issues. For a fixed $\Omega_\textup{T}$ and $V_\textup{0}$,  the Mathieu-$q$ parameter is inversely proportional to the square of the ion-electrode distance $r_\textup{0}$ (see Eq.~\ref{App:Eq7}). Hence $q$ may also be lowered by increasing $r_\textup{0}$. In a three-dimensional linear Paul trap, $r_0$ may be altered by changing the inter-electrode separation, while in a SET, this may be achieved by modifying the electrode widths in the radial plane. Unless special precautions are taken, such as a symmetric rf-drive configuration, this will result in axial micromotion in the transition region. In case of planar traps, owing to the asymmetric rf drive typically used to operate these devices, this micromotion will not be compensated and may hinder ion transport. This modification also leads to an undesired change in the height of the rf null line above the electrode plane, which may be alleviated by a suitable combination of dc potentials~\cite{An_2018}.\\

\noindent Due to its symmetric four-rod geometry, modulating the inter-electrode separation does not alter the position of the ions in a three-dimensional linear Paul trap. The only technical challenge is achieving a precise and smooth modulation of the inter-electrode separation to enable efficient transport of ions from the loading to the experimental zones. The success of 2PP-DLW technique in printing a functional three-dimensional linear Paul trap with $r_0=75~\mu\textup{m}$ promises the requisite fabrication control. The printed trap structures have accuracy and precision $\simeq1~\mu\textup{m}$ \cite{Xu_2025}, which is sufficient to realize the desired structures.\\

\section{Methods}\label{methods}

\begin{figure}[hbt!]
	\includegraphics[width=1\linewidth, left]{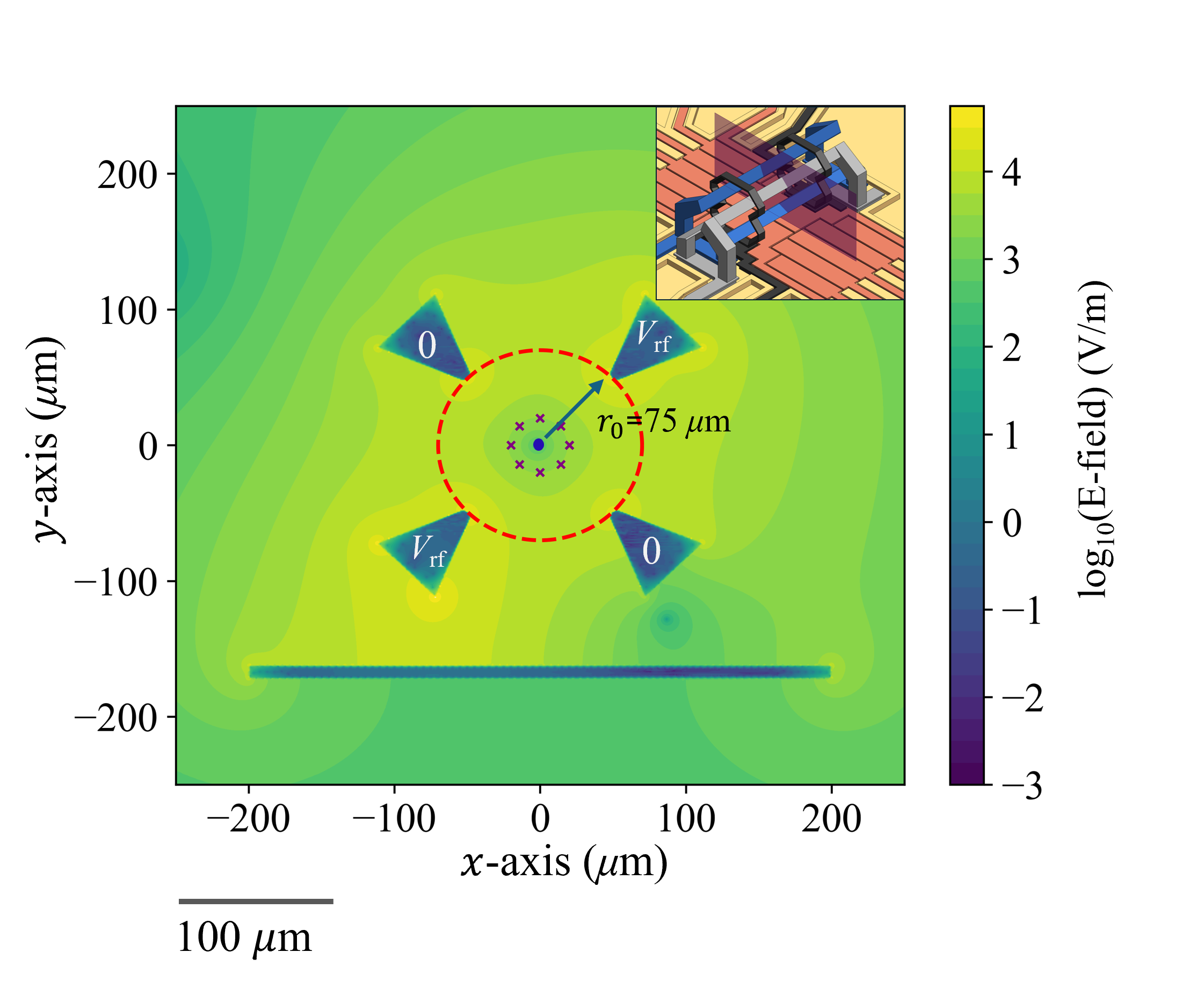} 
	\caption{Radial cross-section of the 3D-printed ion trap used for ion trajectory simulations is shown. The plane (shown in translucent blue in the inset) is located halfway between the end-cap electrodes (in black). An rf voltage is applied to the triangular electrodes. The segmented surface dc electrodes (rf ground) are replaced by a continuous electrode in the simulations. Contours represent the magnitude of electric field for a $1~\textup{V}$ applied voltage. The blue dot represents the minimum of the calculated pseudopotential well. The purple crosses represent the release points for study of trapped fraction for ions created away from the potential minimum (see Sec.~\ref{methods}). \textbf{Inset:} Part of the trap chip is shown. The rf rails are shown in blue and gray (same color represents same polarity). The end-cap electrodes are colored in black. The surface electrodes (copper) are used for micromotion compensation and ion transport operations. The ground surfaces are shown in gold.}
	\label{Fig:1}
\end{figure}

\noindent While reducing the $q$-parameter has been observed to enable faster laser cooling, it also leads to an associated reduction of the trap depth. In this section, we numerically simulate ion trajectories to determine the extent to which the $q$ parameter may be reduced without significantly reducing the number of ions trapped. The classical trajectories of an ensemble of hot ions are simulated for various Mathieu-$q$ parameters to investigate how the number of trapped ions depend on $q$. Ion motion is studied only in the radial plane~(see Fig.~\ref{Fig:1}), as poor laser cooling at high $q$ is induced by strong anharmonicity of the potential and heating~\cite{vanMourik_2022} due to the rf confinement. The axial motion of the ions is ignored, as the dc potentials relevant for that degree-of-freedom can be adjusted during operation.\\

\noindent 
The trap design~\cite{Xu_2025} used here is shown in the inset of Fig.~\ref{Fig:1}. In the simulations, the trap is operated asymmetrically at a drive frequency $\Omega_\textup{T}/{2\pi} = 50~\textup{MHz}$, close to the $\sim51~\textup{MHz}$ used in Ref.~[\onlinecite{Xu_2025}] to achieve radial motional frequencies in excess of $20~\textup{MHz}$. While the pseudopotential approximation may be used to describe motion of cold ions with small excursions from the minimum, it is not adequate for hot ions. Hence the full time-dependent rf potential created by the trap electrodes is used for simulating the ion motion. For this purpose, the AC/DC and the particle tracing modules of COMSOL Multiphysics\textsuperscript\textregistered are used to generate the rf electric potential of the trap and simulate the ion trajectories, respectively.\\

\noindent A resistively-heated source typically produces thermal atoms with a temperature  between $500$ and $1000~\textup{K}$. When an ion is produced in the trap by photoionization, it is at least as hot as the source atom. The laser cooling timescale of hot ions is a few orders of magnitude longer than one period of $\Omega_\textup{T}$ ($20~\textup{ns}$). Therefore, for an ion to be laser-cooled, first it has to be confined in the pseudopotential well of the trap. To isolate the effect of the trapping potential, a simulation time short compared to the timescale of laser cooling is used. An ion is considered trapped if it maintains a bounded trajectory for 2500 periods of the trap rf frequency ($50~\mu\textup{s}$) without laser cooling. Both the neutral atom and the photoionization laser beams have finite sizes. Consequently, ions are not only created at the minimum of the trap potential, but also away from it. However, to keep the problem tractable for numerical simulations, we assume that the ions are created at the minimum of the pseudopotential. The spatial effect of ion creation will be discussed later in this section. An ensemble of $N_\textup{init}=10,000$ $^{40}\textup{Ca}^+$ ions, sampled from a Maxwell-Boltzmann (MB) velocity distribution at a temperature of $1000~\textup{K}$ is used. Here, the initial motion of the ions is perpendicular to the trap axis. Ion-ion interactions are neglected, since the motivation behind using an ensemble is to capture the effect of phase of the driving rf field on the ions' trajectories, and not to study the space-charge induced limitations of loading a trap. After a simulation time $\tau_\textup{sim} = 50~\mu\textup{s}$, the final positions $r_\textup{final} = (x_\textup{f}, y_\textup{f})$ of the ions are recorded. The confined ions are the ones for which $r_\textup{final}$ is within the trapping region~(see Fig.~\ref{Fig:1}a). The fraction of ions trapped,  $f_\textup{trapped}=N_\textup{final}/N_\textup{init}$, is then calculated, where $N_\textup{final}$ is the number of trapped ions at the end of the simulation.\\

\begin{figure}[hbt!]
    \centering
    \includegraphics[width=1.0\linewidth]{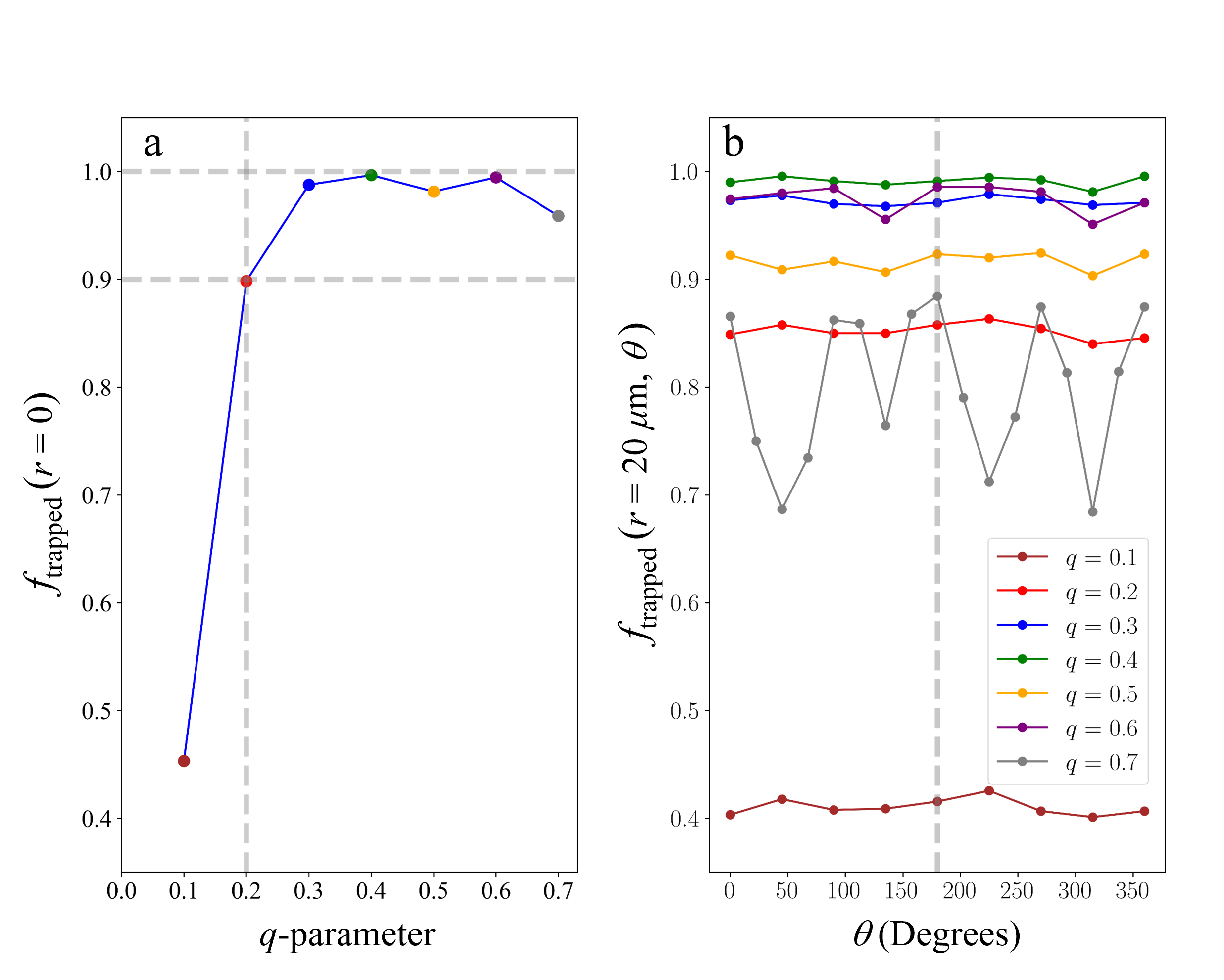}
    \caption{Fraction of ions trapped after $\tau_\textup{sim}=50~\mu\textup{s}$ plotted as a function of Mathieu-$q$ parameter for initial ion temperature of $1000~\textup{K}$. a. When ions are released at the minimum of the trap, no significant deterioration in trapped fraction is observed for $0.2\leq q \leq 0.7$. b. When ions are created $20~\mu\textup{m}$ away from the potential minimum, although absolute number of trapped ions have reduced slightly, $f_\textup{trapped}$ is observed to exhibit minimal dependence on $\theta$ for $q$ between $0.2$ and $0.6$.}
    \label{Fig:2}
\end{figure}

\noindent The trajectory simulations reveal that the trapped fraction remains high, at 0.9 or above, over a wide range of the Mathieu-$q$ parameter (Fig.~\ref{Fig:2}a). In Fig.~\ref{Fig:2}, $f_\textup{trapped}$ is plotted as a function of the Mathieu-$q$ parameter values in the range of $0.1-0.7$ in steps of $0.1$. For $q$ between $0.3$ and $0.7$, the fraction of ions trapped remains practically constant. On the lower end, $f_\textup{trapped}$ drops to $\approx50\%$ for $q=0.1$. The simulations indicate that $\approx90\%$ of the simulated flux from a thermal source is trapped at $q=0.2$, a typical parameter for ion-trapping experiments with macroscopic three-dimensional crystals of cold ions where laser cooling should be efficient. Lowering the $q$-parameter reduces the trapped fraction, while increasing it may decrease the efficiency of laser cooling. While $f_\textup{trapped}$ is only plotted for $50~\mu\textup{s}$, they were also tracked for a range of interim $\tau_\textup{sim}$ to ensure that the curve of $f_\textup{trapped}$ as a function of $q$ is monotonic.\\

\noindent To investigate the effect of the spatial dependence of ion creation on trapped fraction, we assume a photoionization beam diameter of $30~\mu\textup{m}$. Accounting for experimental imperfections, we consider that an ion may be created in a circular region of diameter $40~\mu\textup{m}$ centered at the potential minimum. We simulated the trajectory of 900 ions at $1000~\textup{K}$ released from various points on the aforementioned circle of radius $20~\mu\textup{m}$ for $q$ between $0.1$ and $0.7$. The initial points of release are parameterized with $\theta$, the angle the radius vector $\Vec{r}$ and the $x-\textup{axis}$ (horizontal) subtends at the center of the circle to account for the effects of asymmetry of the potential well. The trapped fraction is plotted as a function of $\theta$ in Fig.~\ref{Fig:2}b. While a minor deterioration in the absolute number of trapped ions is observed when compared with Fig.~\ref{Fig:2}a, the robustness of loading in the three-dimensional geometry is clearly evident for $0.2\leq q\leq 0.6$. At $q = 0.7$, a larger oscillation in trapped fraction is observed with $\theta$.\\

\section{Design and fabrication}\label{fab}

\noindent One of the major advantages of working with a three-dimensional trap with ion-electrode separation $\sim100~\mu\textup{m}$ is that higher secular frequencies may be achieved compared to planar traps. Stable trapping at Mathieu-$q$ parameter $\approx0.9$, corresponding to a $\omega_\textup{rad}=2\pi\times24.15~\textup{MHz}$ has been demonstrated\cite{Xu_2025}. Here, we assume a radial secular frequency of $10~\textup{MHz}$ in the trap in Fig.~\ref{Fig:1}. For a single $^{40}\textup{Ca}^+$ ion, this is achieved at $q=0.57$ for $\Omega_\textup{T}=2\pi\times50~\textup{MHz}$. For a ground-state cooled ion with well-compensated excess micro motion, this operational parameter is well suited, but not for laser cooling of an ion produced either from a thermal oven or via laser ablation.\\

\begin{figure}[hbt!]
	\centering
     \includegraphics[width=1\linewidth]{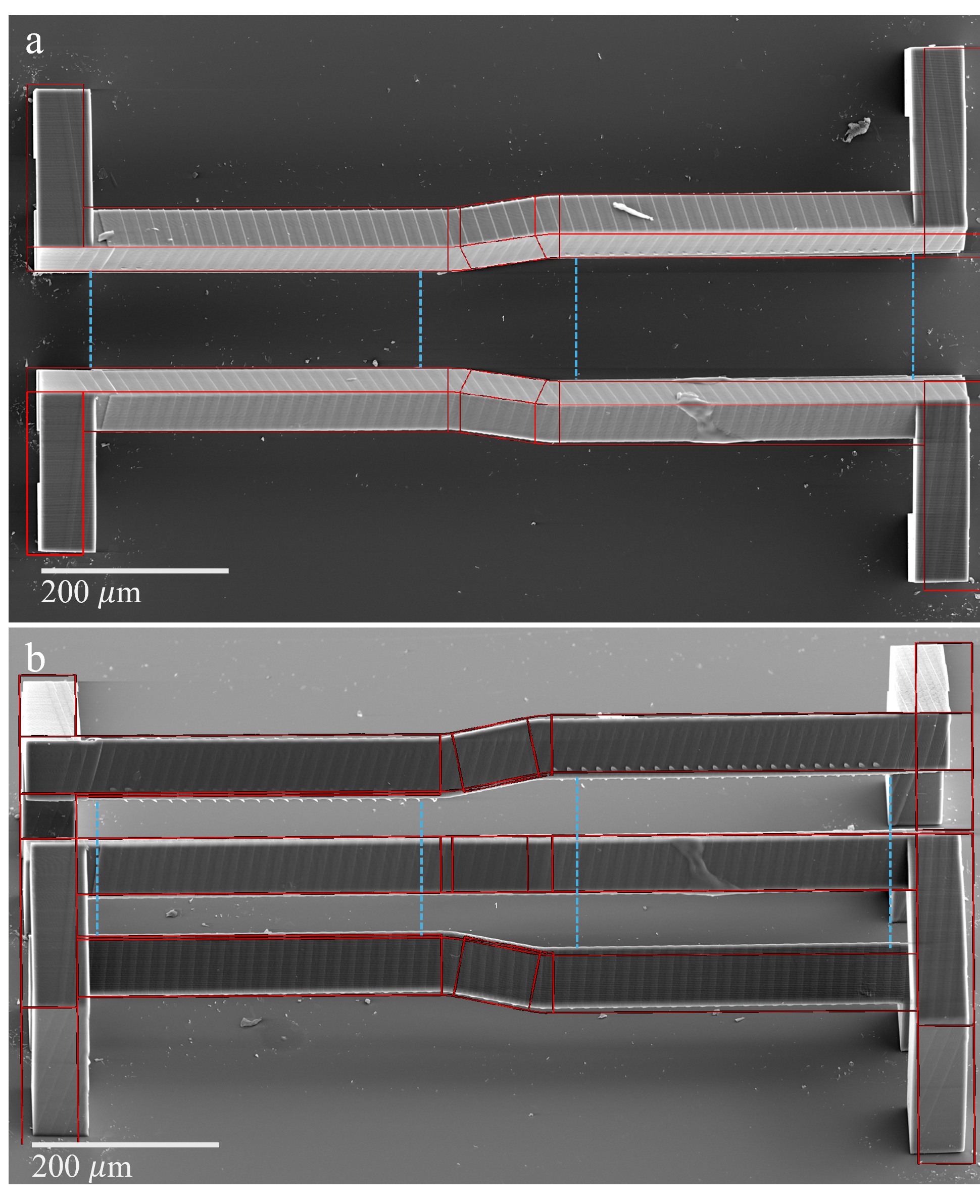}
		\caption{Scanning Electron Microscope images of the fabricated trap rf-electrodes are shown with overlaid CAD references (red) at (a) 0$\degree$ and (b) 45$\degree$. In-plane, the distance between the electrodes measured at the dashed blue lines agree with the design within a few micrometers, while the structure's width is reduced by 0.8\%. Normal dimensions are 5\% smaller than the designed dimensions.}
	\label{Fig:3}
\end{figure}

\noindent To achieve efficient laser cooling of hot ions in the loading zone, we aim to minimize the Mathieu-$q$ parameter while preserving the trapped fraction as much as possible. Based on the simulations in Sec.~\ref{methods}, we target $q\geq 0.2$ in the loading zone while simultaneously supporting high $q$ in the experimental zone in our design. The maximum height of printed structures with the requisite $\simeq1~\mu\textup{m}$ accuracy is $300~\mu\textup{m}$ in our printer. To conform to this limit, $r_0$ is increased from $75~\mu\textup{m}$ to $95~\mu\textup{m}$ corresponding to a change of $q$ from $0.57$ to $0.35$. Since the pseudopotential approximation is assumed to be valid for $q<0.4$, this expansion should lead to more efficient laser cooling. The length of the printed electrodes~(Fig.~\ref{Fig:3}) is $1.02~\textup{mm}$, which may be further extended using supports. The length over which $r_0$ increases is designed to be $\approx120~\mu\textup{m}$, which leads to an opening angle of $\approx10\degree$ in the funnel shaped transition region.\\

\noindent 

\noindent The coupling of the radial and axial potentials in the transition region leads to a pseudopotential barrier when transporting ions from the loading to the experimental region. In simulation, a $^{40}\textup{Ca}^+$ ion at a temperature of $0.5~\textup{mK}$ overcame this potential barrier with an applied voltage of $1~\textup{V}$ (a field strength of $E_\textup{ax}\approx10^3~\textup{V/m}$). If the trap is operated with a symmetric rf drive, this axial barrier vanishes~\cite{An_2018}. In either scenario, adiabatic transport would be sufficient to move ions from the loading zone into the experimental zone.\\

\noindent Only the rf-electrodes are printed for this demonstration. A full device will necessitate either 3D-printing the dc electrodes on the surface~\cite{Xu_2025} or using a chip with pre-fabricated surface electrodes. After fabrication, the inter-electrode distances are measured to assess dimensional accuracy. The printed structure is shown in Fig.~\ref{Fig:3} along with a dimensional comparison with the design. For this measurement, multiple images of the device were taken using a Scanning Electron Microscope and analyzed using ImageJ software\cite{Schneider_2012}. The in-plane distance between the electrodes is in good agreement with the design dimensions; the width of the  structure is reduced by 0.8\%. A small angle ($\approx0.3\degree$) is observed between the straight sections of the same electrode in the experimental and loading zones. The normal distance between the electrodes is 5\% smaller than the designed dimensions (see App.~\ref{App:E}). The discrepancy of the fabricated device compared to the designed one may be reduced by process optimization. However at the current discrepancy level, this leads to a $10\%$ change in the Mathieu-$q$ parameter, requiring a one-time calibration of the applied voltage for each device during commissioning. \\

\section{Discussion and Conclusion}\label{conclusion}

\begin{figure}[hbt!]
	\centering
	\includegraphics[width=0.75\linewidth]{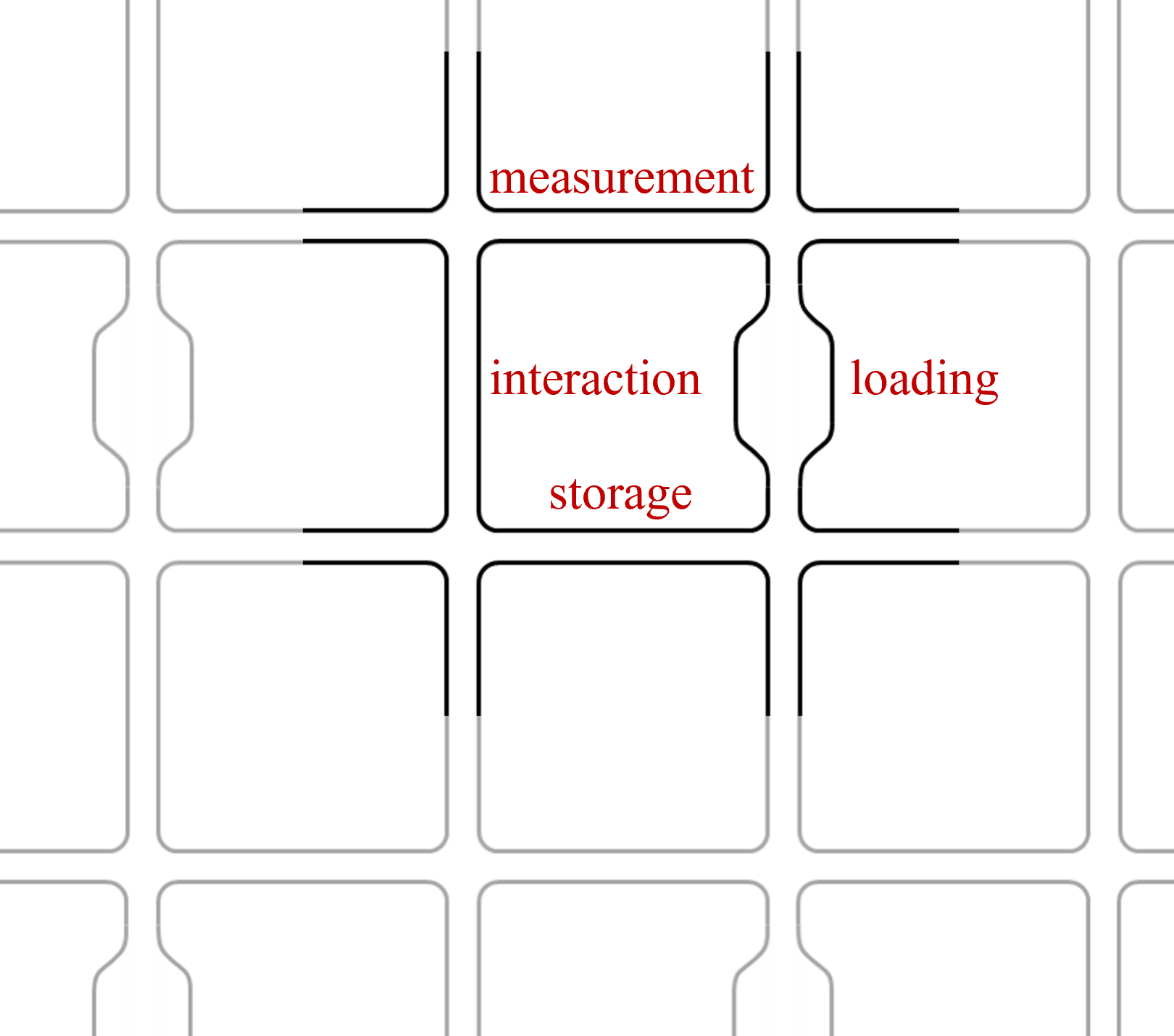}
	\caption{Cartoon of a two-dimensional array of ion traps in a quantum CCD device. Specialized loading regions are shown alongside other functional modules.}
	\label{Fig:4}
\end{figure}

\noindent The precise control of the trap dimensions demonstrated here builds on Ref.~[\onlinecite{Xu_2025}] to demonstrate novel, useful geometries for quantum computing that are realizable by using 3D printing to decorate planar ion traps\cite{Taniguchi_2025}. While existing devices typically have a dedicated loading zone, this fabrication method can generate local geometric modifications that optimize the rf confinement for loading, improving the functionality of this module and potentially reducing one of the major ancillary experimental overheads. These specialized loading modules may be strategically interspersed in an array of traps constituting a large-scale quantum device, such as elements of a repeated tile (Fig.~\ref{Fig:4}) in a two-dimensional array of traps in a QCCD architecture.\\

\begin{figure}[hbt!]
	\centering
     \includegraphics[width=1\linewidth]{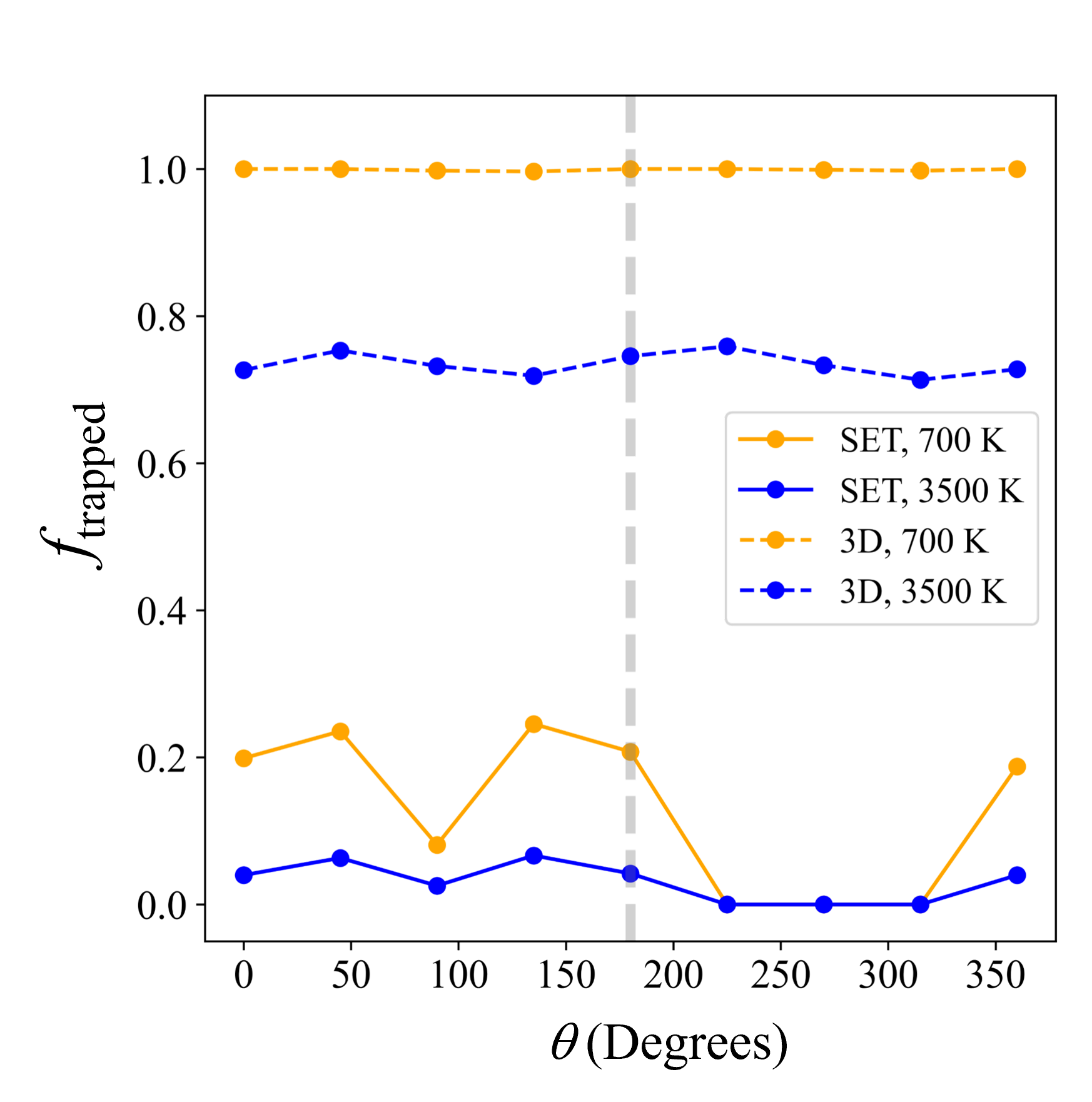}
		\caption{Trapped fraction is plotted as a function of $\theta$ at ion temperatures $700~\textup{K}$ (orange) and $3500~\textup{K}$ (blue) for the three-dimensional trap (dashed line) in Fig.~\ref{Fig:1} and the representative planar trap~\cite{Sage_2012} described in main text (continuous line). The ions are created at a distance of $20~\mu\textup{m}$ from the potential minimum. It is clearly observed that fraction of ions trapped in the planar trap is significantly lower compared to the three-dimensional one. The effect becomes much more pronounced for ablated ions. The variation of the trapped faction with $\theta$ is also more significant in the surface-electrode trap than in the three-dimensional trap.}
	\label{Fig:5}
\end{figure}

\noindent In the context of loading ions, it was mentioned in Sec.~\ref{motivation} that a surface electrode trap operated with parameters similar to a three-dimensional trap may not capture hot ions as efficiently due to strong anharmonicity of the rf potential and shallower associated pseudopotential depth. To support this assertion, simulated trapped fraction in the three-dimensional trap described here is compared with that of a representative surface trap~\cite{Sage_2012} (see Fig.~\ref{Fig:AppC_SET} for a radial cross-section). Both the traps are assumed to operate at $\Omega_\textup{T}/{2\pi} = 50~\textup{MHz}$ with asymmetric voltage drives. The three-dimensional trap is assumed to operate at $q = 0.35$, as achieved in the fabricated loading zone demonstrated in the previous section. The voltage used to drive the surface electrode trap is chosen such that the radial motional frequencies of a single $^{40}\textup{Ca}^+$ ion are equal in both the traps. Numerical simulations predict that at $700~\textup{K}$, a temperature typical for loading of $^{40}\textup{Ca}^+$ ions from thermal oven, fraction of ions trapped in the three-dimensional trap is approximately twice that in the planar trap. For laser ablated ions at $3500~\textup{K}$\cite{Smith_2025}, $f_\textup{trapped}$ is calculated to be more than five times in the three-dimensional trap compared to the surface trap. A comparison of the spatial effect of ion creation between the three-dimensional and the planar trap~(see Fig.~\ref{Fig:AppC_SET} for ion release points) highlights a more striking difference. Fig.~\ref{Fig:5} shows a strong $\theta$ dependence of $f_\textup{trapped}$ for the planar trap when compared with the three-dimensional one, which shows one instance in which loading is more inefficient in a surface electrode trap.\\

\noindent Besides initial loading of ions, owing to the deeper pseudopotential and lower $q$, the geometry described in this article may also improve strategies for reloading of ions in a large scale trapped-ion quantum device~\cite{Bruzewicz_2016, Lekitsch_2017}. Occasionally, an ion string heats up due to background gas collisions and transitions into the unstructured ion-cloud phase. In this scenario, the hot string may be transported to an adjacent loading zone for re-crystallization, which may be less resource intensive than discarding the whole string and reloading. In this context, the loading zone may also be considered as a ``re-crystallization zone". Furthermore, a specialized loading region as described here may be a necessity if trapped-ion quantum computers are to be operated at increasingly higher motional frequencies.\\

\noindent The controllable tuning of the ion-electrode separation may find applications in trapped ion quantum applications beyond enabling efficient ion loading. For example, a tapered linear Paul trap has been used to realize a trapped ion-based single-atom heat engine~\cite{Rossnagel_2016} and demonstrate zeptonewton force sensing\cite{Deng_2023} using controllable coupling between the radial and axial motion. A similar device has recently been proposed for creating non-Gaussian quantum states in the context of continuous variable quantum computation\cite{Nikolova_2025}. The precise fabrication control enabled by 3D-printing may support miniaturization and optimization of trap designs for these applications.\\

\section{Acknowledgements}
\noindent The authors thank Shuqi Xu for providing the trap design used in the simulations. This work was performed under the auspices of the U.S. Department of Energy by Lawrence Livermore National Laboratory under Contract DE-AC52-07NA27344 and was supported by the LLNL-LDRD Program under Project no. 23-ERD-021. HH acknowledges funding from Army Research Laboratory under Award no. W911NF2510233. LLNL-JRNL-2012799.

\section{Author declarations}

\subsection{Conflict of interest}

\noindent The authors have no conflicts to disclose.

\subsection{Data availability}

\noindent The data that support the findings of this study are available from the corresponding author(s) upon reasonable request.

\appendix

\section{Relevant Paul trap equations}\label{App:A}

\noindent For a linear Paul trap with ion-electrode distance $r_0$ operated at a frequency $\Omega_\textup{T}$, the time-dependent potential $U_\textup{td}$ is expressed as

\begin{equation}\label{App:Eq1}
	U_\textup{td} = \frac{V_0\textup{cos}(\Omega_\textup{T}t+\phi)}{2r_0^2}(x^2-y^2).
\end{equation}

\noindent For an ion of mass $m_\textup{ion}$, charge $Z_\textup{ion}\left|e\right|$ trapped in this potential, the equations of motion are given by

\begin{equation}\label{App:Eq2}
	\ddot r = -\frac{Z_\textup{ion}\left|e\right|}{m_\textup{ion}}\frac{\partial U_\textup{td}}{\partial r}.
\end{equation}

\noindent Simplification of Eq.~\ref{App:Eq2} leads to the well known Mathieu equation

\begin{equation}\label{App:Eq3}
	\dv[2]{r}{\xi} - 2q\textup{cos}(2\xi) r = 0,
\end{equation}\\

\noindent where $q$ is the Mathieu-$q$ parameter, defined as

\begin{equation}\label{App:Eq3a}
	q = 2Z_\textup{ion}\left|e\right|V_0/m_\textup{ion}r_0^2\Omega_\textup{T}^2.
\end{equation}\\

\noindent To the lowest order approximation, when $q \ll 1$, the motion of the ion may be described by a pseudopotential $U_\textup{pseudoPE}$ given by

\begin{equation}\label{App:Eq4}
	U_\textup{pseudoPE} = \frac{1}{4m_\textup{ion}\Omega_\textup{T}^2}(\nabla U_\textup{td})^2.
\end{equation}

\noindent In this approximation, the secular frequency of the ion $\omega$ is expressed as

\begin{subequations}\label{App:Eq5}
	\begin{align}
		\omega &= \beta\times \Omega_\textup{T}/2,    \label{App:Eq5a} \\
		\beta &= q/\sqrt{2}, \;q^2 \ll 1.   \label{App:Eq5b}
	\end{align}				
\end{subequations}

\noindent Hence,

\begin{equation}\label{App:Eq6}
	\omega = q\times\frac{\Omega_\textup{T}}{2\sqrt{2}}.
\end{equation}\\

\noindent The operational rf voltage $V_{0, \textup{asym}}$ for a linear Paul trap driven asymmetrically is given by
\begin{equation}\label{App:Eq7}
	V_{0, \textup{asym}} = q\frac{m_\textup{ion} r_\textup{0}^2 \Omega_\textup{T}^2}{2 Z_\textup{ion}\left|e\right|}.
\end{equation}

\noindent If the trap is operated with symmetrical voltage drive, $V_{0, \textup{sym}}$ is given by
\begin{equation}\label{App:Eq8}
	V_{0, \textup{sym}} = q\frac{m_\textup{ion} r_\textup{0}^2 \Omega_\textup{T}^2}{4 Z_\textup{ion}\left|e\right|}.
\end{equation}

\section{Ion trajectory simulations}\label{App:C}

\noindent To generate the potential for ion trajectory simulations, finite element method (FEM) is used. To create a smooth potential, the simulation domain is meshed with elements of minimum and maximum size $0.01$ and $5~\mu\textup{m}$, respectively. The particle trajectories were saved ten times in each period ($20~\textup{ns}$) of the trap rf frequency of $\Omega_\textup{T}/{2\pi} = 50~\textup{MHz}$, while the actual time steps used in the simulations were much smaller. As explained in the main text, a simulation time of $50~\mu\textup{s}$ is used. The ions are created at the minimum of the pseudopotential, and released at time $t = 0$. The effect of phase of the rf potential on trapped fraction is captured by using a large ensemble of $10,000$ ions sampled from a MB distribution at $1000~\textup{K}$.\\

\begin{figure}[hbt!]
	\centering
     \includegraphics[width=1\linewidth]{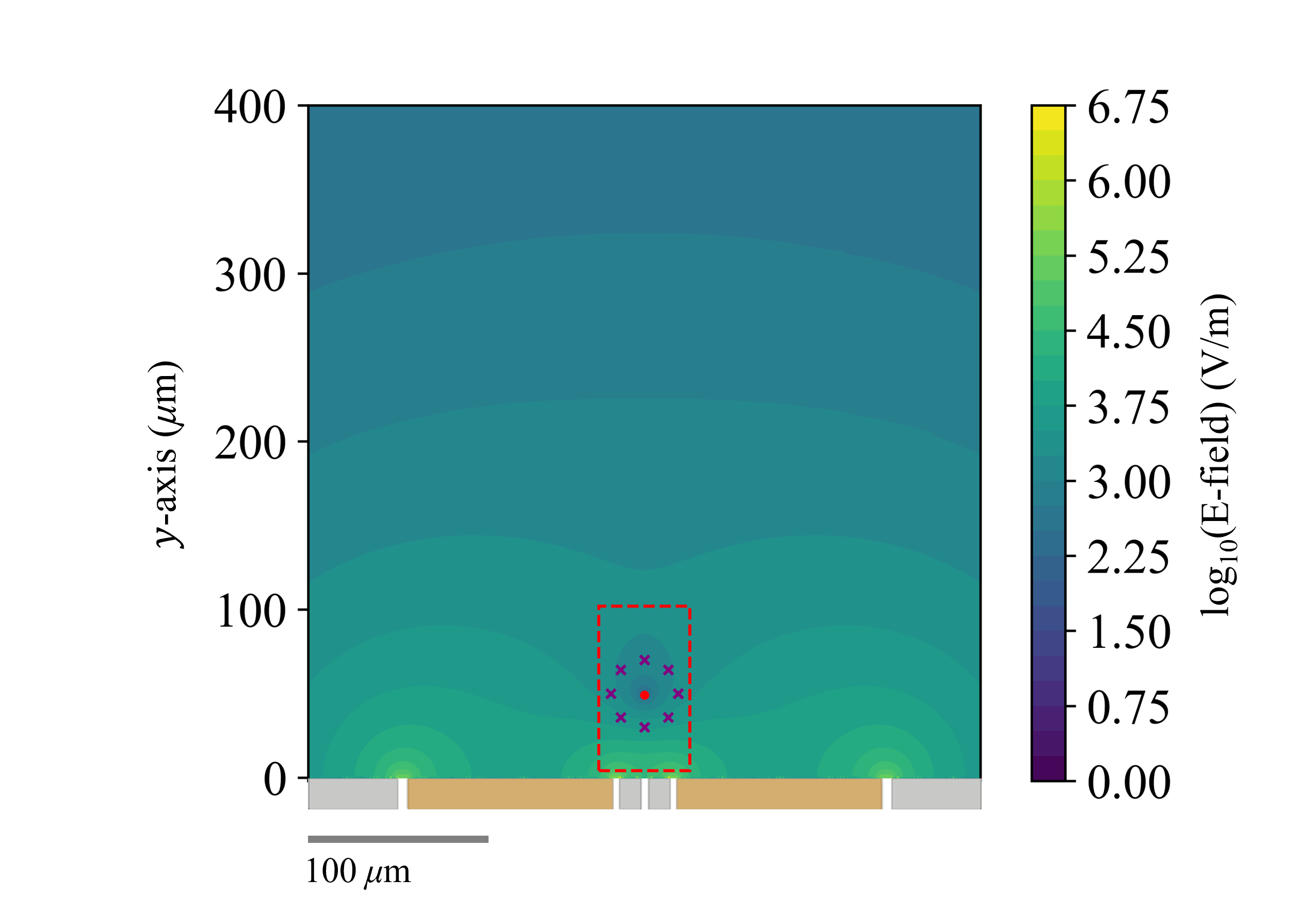}
    \caption{A cross-section of the planar trap used for ion trajectory simulations. The rf voltage is applied to the electrodes in gold, while the gray ones are held at ground. The contours represent the magnitude of electric field for a $1~\textup{V}$ applied voltage. The red dot represents the calculated minimum of the pseudopotential well. The trapping region as described in the text is represented by the dashed rectangle. The release points of ions for simulating the trapped fraction when ions are created $20~\mu\textup{m}$ away from the potential minimum is shown by the purple crosses.}
	\label{Fig:AppC_SET}
\end{figure}

\noindent For simulating the trajectories of ions created $20~\mu\textup{m}$ away from the potential minimum, an ensemble of 900 ions sampled from the same MB distribution at $1000~\textup{K}$ was used to reduce the total computational time. Release points at intervals of $\pi/4$, starting with $\theta = 0$ were used (orange crosses in Fig.~\ref{Fig:AppC_SET} for the SET). A denser grid of release points at intervals of $\pi/8$ were used for $q = 0.7$ in the three-dimensional trap to investigate the larger oscillation of $f_\textup{trapped}$ with $\theta$.\\

\noindent The radial frequency of a single $^{40}\textup{Ca}^+$ ion in the three-dimensional trap at $q = 0.35$ is $\omega_\textup{rad}/2\pi = 5.5~\textup{MHz}$. To extract $\omega_\textup{rad}$, trajectory of a single $^{40}\textup{Ca}^+$ ion at $1~\textup{mK}$ is simulated and Fourier transformed. To find the appropriate operating voltage for the planar trap, trajectories of the same ion is simulated for a range of rf voltages. An rf voltage of $47.5~\textup{V}$ is found to achieve $\omega_\textup{rad}/2\pi \approx 5.5~\textup{MHz}$.

\noindent The trapping region in the three-dimensional trap is defined by a circle of $r_0$~(see Fig.~\ref{Fig:1}). This circle traces out the physical barrier constituted by the electrodes and the turning points in the pseudopotential well in both $x$ and $y$-directions. For the planar trap, to define the trapping region, the turning points of the pseudopotential well in directions parallel ($x$-axis) and perpendicular ($y$-axis) to the surface electrodes were first calculated. These are represented by ($x_\textup{min}$, $x_\textup{max}$) and ($y_\textup{min}$, $y_\textup{max}$) for the $x$ and $y$-axes respectively. The trapping region is then determined by a rectangle with sides $x_\textup{max}-x_\textup{min}$ and $y_\textup{max}-y_\textup{min}$~(see Fig.~\ref{Fig:AppC_SET}). In case of the planar trap, this is an overestimate of the trapping region, and the actual trapping region is smaller than what is calculated here.\\

\section{3D-printing of the trap}\label{App:D}

\noindent The structures were printed using Nanoscribe GT2 2PP printer using IP-S resin and x25 objective. The CAD files were processed using Nanoscribe's Describe software and the parts were sliced using a method defined in Ref.~[\onlinecite{Marschner_2023}]. Briefly, the printed structures were sliced with 0.8 micron spacing and hatching was set to be $0.4~\mu\textup{m}$ spacing. Each individual rf beam was separated into its own part starting from the lower and then the top rf current carrying beams. Each individual beam was sliced into $16~\mu\textup{m}$ blocks with an angle of $15\degree$ and an overlap of $8~\mu\textup{m}$ with each block. After printing, the structures were developed using PGMEA and isopropyl alcohol solutions, dried under compressed air, and then sputter coated with gold with a thickness of $100~\textup{nm}$ for imaging. Scanning electron microscopy (SEM FEI Apreo) was used for imaging the printed structures.\\

\section{Printed structure dimensions}\label{App:E}

\noindent After extracting and applying the SEM scale and rotating images so that the upper left rf rail was horizontal, ImageJ's built-in measuring tool was used to extract the dimension between the top upper electrode and lower bottom electrode at four locations (see Fig.~\ref{Fig:3}). Additionally, the width of the trap was measured at the top and bottom of the supports. A summary of the extracted information can be found in Tab.~\ref{tab:my_label}.

\begin{table}[hbt!]
    \centering
    
    \begin{tabular}{|l|c|c|c|} \hline 
          \textbf{Location}&\textbf{Angle}&\textbf{Expected}& \textbf{Measured}\\ \hline 
          Electrode separation:&0$\degree$&102.3&105.6 $\pm$ 1.6\\ 
          Experimental zone&45$\degree$&144.7&144.2 $\pm$ 0.7\\ 
          &90$\degree$&102.3&98.3 $\pm$ 2.6\\ \hline 
          Electrode separation:&0$\degree$&129.83&134.5 $\pm$ 1.1\\ 
          Loading Zone&45$\degree$&183.62&180.4 $\pm$ 0.3\\ 
          &90$\degree$&129.83&120.6 $\pm$ 1.6\\ \hline 
          Trap Width&0$\degree$&1003.28& 996.0 $\pm$ 3.7\\ 
          &45$\degree$&1003.28&994.0 $\pm$ 4.4\\ \hline
    \end{tabular}
    
    \caption{Extracted dimensions in $\mu$m of the printed trap from SEM images. 3 images were included in the analysis for 0$\degree$; 2 were included at 45$\degree$. Dimensions at 90$\degree$ are derived from measurements at 0$\degree$ and 45$\degree$.} 
    \label{tab:my_label}
\end{table}

\bibliography{Draft_main_text_05062026_rev_13}

\end{document}